# Self-Consistent Computation of Spin Torques and Magneto-Resistance in Tunnel Junctions and Magnetic Read-Heads with Metallic Pinhole Defects

Serban Lepadatu[1,*], Alexey Dobrynin[2]

[1]*Jeremiah Horrocks Institute for Mathematics, Physics and Astronomy, University of Central Lancashire, Preston PR1 2HE, U.K.*
[2]*Seagate Technology, 1 Disc Drive, Derry, BT48 0BF, U.K.*

**Abstract**

A three-dimensional self-consistent spin transport model is developed, which includes both tunnelling transport, leading to tunnelling magneto-resistance, as well as metallic transport, leading to giant magneto-resistance. Using the spin accumulation computed either side of a tunnel barrier, spin torques are obtained, and it is shown the model reproduces both damping-like and field-like spin-transfer torques, with the expected sinusoidal angular dependence, and inverse ferromagnetic layer thickness dependence. An explicit solution to the drift-diffusion model is derived, which allows analysing the effect of both the reference and free layer thickness on the spin-transfer torque polarization and field-like coefficient. In particular, when the layers are thin, additional spin-dependent scattering contributions due to incomplete absorption of transverse spin components reduce both the damping-like and field-like spin torques. It is shown the model developed here can be used to compute the signal-to-noise ratio in realistic magnetic read-heads, where spin torque-induced fluctuations and instabilities limit the maximum operating voltage. The effect of metallic pinhole defects in the insulator layer is also analysed, which results in a mixture of metallic and tunnelling transport, and highly non-uniform charge and spin currents, requiring the full spin transport model to compute the resulting magneto-resistance and spin torques. Increasing the area covered by pinholes results in a rapid degradation of the magneto-resistance, following an inverse dependence. Moreover, the spin torque angular dependence becomes skewed, similar to that obtained in fully metallic spin valves, and the spin-transfer torque polarization decreases. The same results are obtained when considering tunnel junctions with a single pinhole defect, but decreasing cross-sectional area, showing that even a single pinhole defect can significantly degrade the performance of tunnel junctions and magnetic read-heads below the 40 nm node.

[*]SLepadatu@uclan.ac.uk

# I. Introduction

Magnetic tunnel junctions (MTJ) are one of the most important building blocks for a number of spintronics devices, including magnetic read-heads [1,2], spin-transfer torque magnetic random access memory (STT-MRAM) [3], spin torque nano-oscillators [4], and magnetic field sensors [5]. In an MTJ, a thin insulator layer separates two ferromagnetic layers, with resistance strongly dependent on the relative magnetization orientations of the two ferromagnets. This effect is known as tunnelling magneto-resistance (TMR), and is typically an order of magnitude greater than current perpendicular to plane giant magneto-resistance (CPP-GMR), where the two ferromagnets are separated by a metallic spacer. Another effect which occurs both in TMR and CPP-GMR stacks is spin-transfer torque (STT), where a current spin-polarized by one ferromagnet, will exert a spin torque on the other ferromagnet, as transverse spin components of itinerant electrons are transferred to localized lattice electrons. STT can result in magnetization switching [6], which is utilized in STT-MRAM [3], however in magnetic read-heads STT is generally undesirable due to induced instability and increased magnetic noise [7,8]. The typical approach to including STT in computations is with a macrospin approximation [9]. This is sufficient if currents are uniform and the modelled stack comprises only 2 ferromagnetic layers. In general however, a more powerful approach is required, which is able to take into account not only non-uniform charge and spin currents, but also the non-local nature of STT, as arising from spin-dependent scattering at multiple interfaces.

A possible approach to take into account the non-local nature of STT in multi-layered stacks is to employ a drift-diffusion model, as previously applied to MTJs [10]. Numerical drift-diffusion modelling has been previously used for spin torque and CPP-GMR computation in fully metallic spin valves [11]. Previous works have also used a multiscale model to model tunnel junctions, combining atomistic spin dynamics with *ab initio* STTs [12]. Another approach to self-consistently computing spin torques, used with the Landau-Lifshitz-Gilbert (LLG) equation, was demonstrated by using a time-dependent non-equilibrium Green's function algorithm [13]. More recently another method has been proposed to compute STTs, using a matrix-based non-equilibrium Green's function algorithm to couple micromagnetic simulations with ballistic transport in magnetic tunnel junctions [14]. In this work we develop a drift-diffusion model including both tunnelling and metallic transport, which can be applied to model three-dimensional devices with realistic dimensions efficiently, such as magnetic read-heads, including both TMR and CPP-GMR contributions. We further use this model to analyse the effect of metallic pinholes in the insulator layer. Such defects can occur in thin insulator layers, or after dielectric breakdown [15]. Previous works have concentrated on the effects of pinholes on magneto-resistance [16-21]. A self-consistent drift-diffusion model allows analysis not only of magneto-resistance, including TMR and CPP-GMR contributions, but also of spin torques. In particular we show how the sinusoidal angular dependence of STT in MTJs becomes skewed, and the STT efficiency decreases, as the pinhole concentration increases, tending towards the STT characteristics obtained in fully metallic stacks. This is also important in magnetic read-heads, where we show that the presence of even a single pinhole can greatly restrict the operating reading voltage range, owing to increased magnetic noise and STT-induced instabilities.

This work is organized as follows. In the next Section the drift-diffusion model is introduced, applicable to three-dimensional structures including both insulator and metallic spacers between ferromagnetic layers, as well as metallic conduction channels in the insulator layer. In Section III the model is used to analyse spin transport in simple MTJs, comprising 2 ferromagnetic layers, both with and without metallic pinholes in the insulator layer. Such structures need to include backing metallic contacts, unless the ferromagnetic layers are very thick. In this case we show how spin-dependent scattering at the ferromagnet-metal interfaces from incompletely absorbed transverse spin components, results in additional diffusive contributions, with STTs dependent on both the ferromagnet layer thicknesses. We introduce an analytical approximation, termed the orthogonal approximation, which is able to reproduce these additional diffusive contributions. The orthogonal approximation and its limit of applicability is further discussed in Appendix B. Finally, realistic magnetic read-head devices, used as hard disk drive (HDD) readers, are analysed in Section IV, where the non-local nature of spin transport, as arising from spin-dependent scattering at multiple interfaces, is taken into account using the self-consistent drift-diffusion model. Here we show how the model introduced in this work is able to compute the signal-to-noise ratio in HDD TMR read-heads as a function of reading voltage, both with and without metallic pinhole defects.

## II. Model

Charge and spin current densities are calculated using a drift-diffusion model [22-24] as:

$$\mathbf{J}_C = \sigma\mathbf{E} + \beta_D D_e \frac{e}{\mu_B}(\nabla\mathbf{S})\mathbf{m},$$

$$\mathbf{J}_S = -\frac{\mu_B}{e} P\sigma\mathbf{E}\otimes\mathbf{m} - D_e\nabla\mathbf{S}.$$

(1)

Here $\sigma$ is the electrical conductivity, $D_e$ is the electron diffusion constant, $\beta_D$ is the diffusion spin polarization which results in CPP-GMR, $\mathbf{m}$ is the magnetization direction, and $\mathbf{S}$ is the spin accumulation. The spin current density expression includes a drift term, with $P$ the current spin polarization, and a diffusion term due to gradients in the spin accumulation. As usual $e$ is the electric charge and $\mu_B$ the Bohr magneton. In tunnel barriers separating two ferromagnetic layers, such that $\cos\theta$ is the dot product of pairs of magnetization unit vectors either side of the barrier, we calculate the conductivity from the angular-dependent resistance, $R(\theta)$, obtained from Slonczewski's formula [25]:

$$R(\theta) = \frac{R_0}{1 + \frac{R_{ap} - R_p}{R_{ap} + R_p}\cos\theta}.$$

(2)

This relation has been verified experimentally [26], where $R_0 = 2R_{ap}R_p/(R_{ap} + R_p)$ is the resistance for perpendicular orientation of ferromagnetic layers' magnetizations, with $R_{ap}$ and $R_p$ being resistances for antiparallel and parallel orientations respectively. While in Ref. [22] TMR was defined as $TMR = 2(R_{ap} - R_p)/(R_{ap} + R_p)$, here we use another definition: $TMR = (R_{ap} - R_p)/R_p$, which is accepted as standard now [27,28]. The tunnel barrier's resistance area ($RA$) product is obtained for the parallel orientation of FL and RL magnetization vectors, i.e. $RA = R_p A$, and it is used as a parameter in the model. The conductivity is then expressed as $\sigma = d_I/R(\theta)A$, where $d_I$ is the insulator layer thickness and $A$ is the barrier area.

Thus, within this model the insulator layer is treated as a poor conductor, and a charge current density is computed throughout a simulated structure by taking the electrical potential *V* (where $\mathbf{E} = -\nabla V$), to be continuous across all interfaces. Advanced models of computing TMR are available, including *ab initio* modelling and tight-binding theory [29,30], however such models are computationally expensive when applied to realistic devices, and the current approach suffices for modelling of magnetic processes.

For spin transport, the spin accumulation follows the equation of motion [31]:

$$\frac{\partial \mathbf{S}}{\partial t} = -\nabla \cdot \mathbf{J}_S - D_e \left( \frac{\mathbf{S}}{\lambda_{sf}^2} + \frac{\mathbf{S} \times \mathbf{m}}{\lambda_J^2} + \frac{\mathbf{m} \times (\mathbf{S} \times \mathbf{m})}{\lambda_\varphi^2} \right). \tag{3}$$

Here $\lambda_{sf}$ is the spin flip length (or spin diffusion length), $\lambda_J$ is the exchange rotation length, and $\lambda_\varphi$ is the spin dephasing length. Since the spin transport dynamics are typically up to 3 orders of magnitude faster than the magnetization dynamics, Equation (3) is solved in the static limit [32], obtaining the Poisson equations:

$$\nabla^2 V = \frac{\beta_D D_e}{\sigma} \frac{e}{\mu_B} \nabla \cdot (\nabla \mathbf{S}) \mathbf{m},$$

$$\nabla^2 \mathbf{S} = -\frac{\mu_B}{e} \frac{P\sigma}{D_e} (\mathbf{E} \cdot \nabla) \mathbf{m} + \frac{\mu_B}{e} \frac{P\sigma}{D_e} (\nabla^2 V) \mathbf{m} + \frac{\mathbf{S}}{\lambda_{sf}^2} + \frac{\mathbf{S} \times \mathbf{m}}{\lambda_J^2} + \frac{\mathbf{m} \times (\mathbf{S} \times \mathbf{m})}{\lambda_\varphi^2}. \tag{4}$$

Equation (4) is solved throughout the simulated structure by taking the spin accumulation and spin currents to be continuous across all interfaces. Further, across the tunnel barrier we impose the requirement of net spin current conservation, which reduces Equation (4) to Laplace equations within the tunnel barrier – within this model spin currents are conserved across the tunnel barrier, but discontinuities in the spin accumulation can arise between the two sides of the barrier. Moreover, within metallic conduction channels in the tunnel barrier (pinhole defects), the full Equation (4) is solved.

Using the spin accumulation, spin torques are computed self-consistently with Equation (5), which arise due to absorption of transverse spin components (transverse to the magnetization direction), on a length-scale governed primarily by $\lambda_J$ and $\lambda_\varphi$.

$$\mathbf{T}_S = -\frac{D_e}{\lambda_J^2}\mathbf{m}\times\mathbf{S} - \frac{D_e}{\lambda_\varphi^2}\mathbf{m}\times(\mathbf{m}\times\mathbf{S}) \tag{5}$$

It is known that spin torques in MTJs follow a simple sinusoidal dependence on the angle between a fixed layer magnetization **p** and a free layer magnetization **m** [27,33], with damping-like (DL-STT) and field-like (FL-STT) components:

$$\mathbf{T}_{STT} = T_{STT}\mathbf{m}\times(\mathbf{m}\times\mathbf{p}) + \beta T_{STT}(\mathbf{m}\times\mathbf{p}),$$
$$T_{STT} = \frac{\mu_B \eta J_{STT}}{e d_F}. \tag{6}$$

Here $d_F$ is the free layer thickness, $\eta$ is an STT polarization constant, $\beta$ is the field-like coefficient, and $J_{STT}$ is the charge current density normal to the tunnel barrier, defined here for the perpendicular configuration (when **m** and **p** are perpendicular). As we show below, the self-consistent spin torque in Equation (5) reproduces $\mathbf{T}_{STT}$, both in the angular dependence and inverse free layer thickness dependence. For metallic spin valves, the spin torques are also given by Equation (6), with the important distinction that $\eta$ is no longer a constant, but also depends on the angle $\theta$ (the angle between **m** and **p**) as [9]:

$$\eta(\theta) = \frac{\eta_0}{1+\chi\cos\theta} \tag{7}$$

Here $\eta_0$ is a constant STT polarization, and $\chi$ is the STT skew – for metallic spin valves the angular dependence is skewed from the simple sinusoidal dependence found in MTJs [34]. Thus, for MTJs we have $\chi = 0$, but for metallic spin valves $\chi > 0$.

## III. Tunnel Junctions

As a first test, the spin accumulation is computed in a CoFeB (F) / MgO (I) tunnel junction, including non-magnetic metallic contacts (N), as shown in Figure 1(a). Here the MgO barrier has 1 nm thickness, whilst the F and N layers have 30 nm thickness, with square cross-sectional area of 40 nm × 40 nm. Electrodes are defined at the *z*-axis ends of the N contacts, with a 1 mV potential drop set. The computed spin accumulation is shown in Figure 1(b) along the *z* direction, where the left F layer (the reference layer) has magnetization along the *x* direction and the right F layer (the free layer) has magnetization along the -*y* direction. Parameters used, and simulation details, are given in Appendix A.

**Figure 1** – Spin accumulation computed in a CoFeB (F) / MgO (I) tunnel junction with metallic contacts (N). (a) Diagram of tunnel junction structure, with left F layer magnetization along *x* direction and right F layer magnetization along -*y* direction. The *z* direction spin current polarization is shown in the N layers. (b) Computed spin accumulation, shown along the z direction.

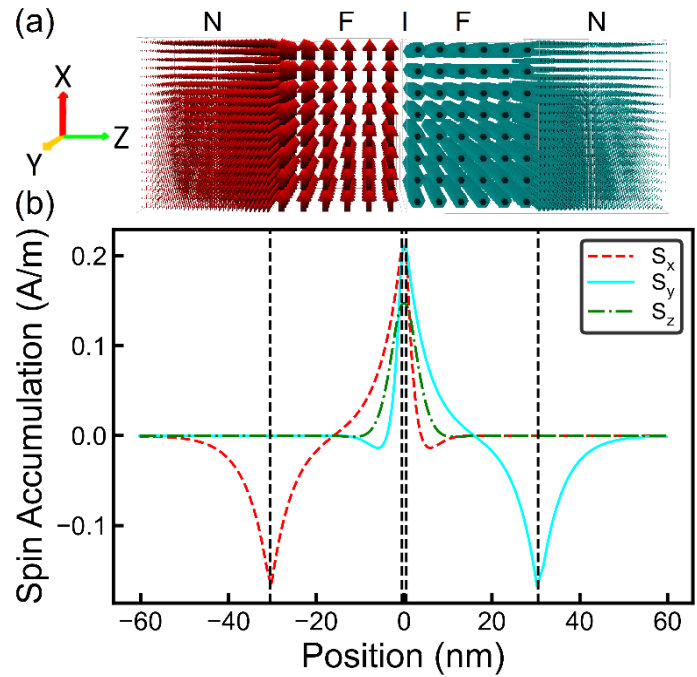

As expected, longitudinal spin accumulation components are generated at the N/F and F/I interfaces, since the spin current is polarized along the local magnetization direction in the F layers. Transverse spin accumulation components are also generated either side of the tunnel barrier, which are absorbed on a short length-scale, giving rise to STT. The absorption of transverse spin components is governed primarily by $\lambda_J$ and $\lambda_\varphi$, and appears as an exponentially decaying oscillation due to the exchange interaction between itinerant and localized electrons [23,31]. This picture is in agreement with tight-binding calculations based on the non-equilibrium Green functions approach [33,35], also revealing on oscillatory dependence of the spin torque with distance from the interface, and large DL-STT and FL-STT components close to the interface. Moreover, the angular dependence of the DL-STT and FL-STT components are plotted in Figure 2(a), showing the expected sinusoidal dependence.

**Figure 2** – Spin torques in an ideal tunnel junction, computed self-consistently using the spin transport solver, with a fixed reference layer thickness of 30 nm and varying free layer thickness. (a) DL-STT and FL-STT as a function of angle with a sinusoidal fit. (b) STT polarization ($\eta$) and field-like coefficient ($\beta$) as a function of free layer thickness obtained numerically, compared to the analytical solution from Equation (9) – the thick F layer approximation – and the orthogonal approximation of Appendix B.

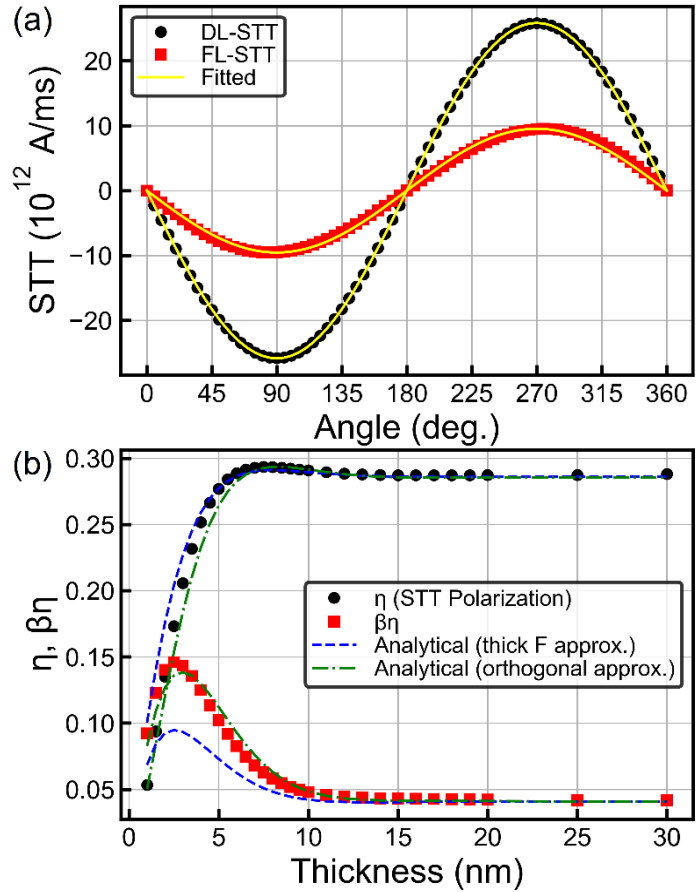

For the simple case of thick F layers (i.e. transverse components fully absorbed) with uniform magnetization it is possible to solve the drift-diffusion model analytically, e.g. as done in Refs. [36-38]. Here we extend the solution to the N/F/I/F/N stack by including the spin dephasing length $\lambda_\varphi$, as well as allowing different transport parameters for the N and F layers. For thin F layers the computations require inclusion of the N layers due to the additional boundary conditions when transverse spin components are not fully absorbed – e.g. in real devices the N layers would be the metallic contacts used to apply a voltage. The extension of analytical solutions to thin F layers is also discussed. With the thick F layer approximation the spin accumulation solution in the right F layer is obtained as:

$$S_x(z) = S_F \exp(-k_1 z)[\cos(k_2 z) - r\sin(k_2 z)]$$
$$S_y(z) = S_F \exp(-z/\lambda_{sf}) + S_N \exp((z - d_F)/\lambda_{sf}), \quad (8)$$
$$S_z(z) = S_F \exp(-k_1 z)[\sin(k_2 z) + r\cos(k_2 z)].$$

Here $(k_1 \pm i k_2)^2 = \lambda_{sf}^{-2} + \lambda_\varphi^{-2} \pm i\lambda_J^{-2}$, $r = k_2/k_1$, $S_F$ and $S_N$ are longitudinal spin accumulation values at the I/F and F/N interfaces respectively. Using continuity of spin currents and spin accumulation at interfaces, the boundary values are obtained as $S_F = -c/D_{e,F}(\lambda_{sf}^{-1} + k)$ and $S_N = c/(D_{e,F}\lambda_{sf}^{-1} + D_{e,N}\lambda_{sd}^{-1})$, where $c = \mu_B P J_{STT}/e$, $\lambda_{sd}$ is the spin diffusion length in the N layer, and $k = k_1 + r k_2$. As we have verified, Equation (8) reproduces accurately the computed spin accumulation in Figure 1(b). Finally, the spin torque is obtained by substituting Equation (8) in Equation (5), and comparison with Equation (6) obtains the thickness-dependent STT polarization and field-like coefficients as:

$$\eta(d_F) = P \int_0^{d_F} \exp(-k_1 z)[k_a \cos(k_2 z) - k_b \sin(k_2 z)]dz,$$

$$\beta(d_F) = \frac{\int_0^{d_F} \exp(-k_1 z)[b\cos(k_2 z) + a\sin(k_2 z)]dz}{\int_0^{d_F} \exp(-k_1 z)[b\sin(k_2 z) - a\cos(k_2 z)]dz}. \quad (9)$$

Here we have $k_a = (r\lambda_J^{-2} + \lambda_\varphi^{-2})/(\lambda_{sf}^{-1} + k)$, and $k_b = (r\lambda_\varphi^{-2} - \lambda_J^{-2})/(\lambda_{sf}^{-1} + k)$. Solving the integrals in the limit of thick F layers we obtain:

$$\eta = P \frac{k_a k_1 - k_b k_2}{k_1^2 + k_2^2},$$

$$\beta = \frac{k_b k_1 + k_a k_2}{k_b k_2 - k_a k_1}.$$

(10)

It should be noted the STT polarization and field-like coefficients depend on both the reference and free layer thickness, since for thin F layers the transverse spin components are not fully absorbed, and instead penetrate through to the backing N layer. Thus, in the limit of small $d_F$ the STT polarization tends to zero, and also the field-like coefficient increases, reflecting the large FL-STT contribution close to the tunnel barrier [35]. This is shown by the numerical results in Figure 2(b), where the reference layer has 30 nm thickness, and free layer has variable thickness, obtained by fitting $\eta$ and $\beta$ to the spin torque from Equation (5). Here, the analytical results from Equation (9) are also given, showing a good agreement with the numerical results in the limit of large $d_F$. For small $d_F$ values the analytical results become inaccurate due to the simplifying assumptions used to derive them. In particular, since significant transverse spin current components are incident on the F/N interface, additional exponential spin accumulation terms are generated in the F layer for the transverse components. These are diffusive contributions to the total spin current caused by spin-dependent scattering. As shown in Figure 2(b) the thick F layer approximation can result in inaccurate spin torque parameters, particularly for the FL-STT component. Moreover, the results in Figure 2(b) are obtained with a thick reference layer (30 nm). As shown in Appendix B, if both the reference and free layers have realistic thickness values, e.g. as in Ref. [38] where Equation (8) was used, the approximation of thick F layers becomes unsuitable and the additional spin-dependent scattering contributions on both sides of the barrier must be taken into account. As discussed in Appendix B, in the limit of thin F layers the equations become implicit, and a numerical method is required, or alternatively a better approximation can be used to allow an explicit solution. Such an approximation may be obtained by including the additional transverse exponential decay terms, but ignoring the resulting coupling between the $k_1$, $k_2$ terms and boundary values. We call this the orthogonal approximation, which is discussed in detail in Appendix B. Results are shown in Figure 2(b), and as may be observed the numerical results are closely reproduced even for thin free layer. In general, the numerical approach is more powerful since it's also applicable to F layers with non-uniform magnetization, and moreover if pinhole defects are to be considered, the resulting non-uniform charge and spin currents require a full numerical method to resolve the non-local nature of the resulting spin torques. As noted in the introduction, a number of methods exist for self-consistent computation of spin torques, however the drift-diffusion model remains a good compromise between *ab initio* modelling at one extreme, and macrospin approximation at the other extreme.

**Figure 3** – Magneto-resistance in a tunnel junction with metallic pinhole defects, computed through the whole N/F/I/F/N stack using the spin transport solver. (a) Angular dependence of resistance, shown for varying pinhole percentage (percentage of area covered by pinhole defects), including the ideal case with no defects (TMR). As a limiting case, the result obtained for a metallic spin valve is also shown (CPP-GMR). (b) Degradation of magneto-resistance with increasing pinhole percentage. Solid symbols are for a fixed square cross-sectional area of 40 nm × 40 nm and with multiple pinholes, whilst open symbols are for a single pinhole but decreasing cross-sectional area. The inset illustrates pinhole defects in the insulator layer, showing the effective conductivity, ranging from tunnelling (blue), to metallic (red).

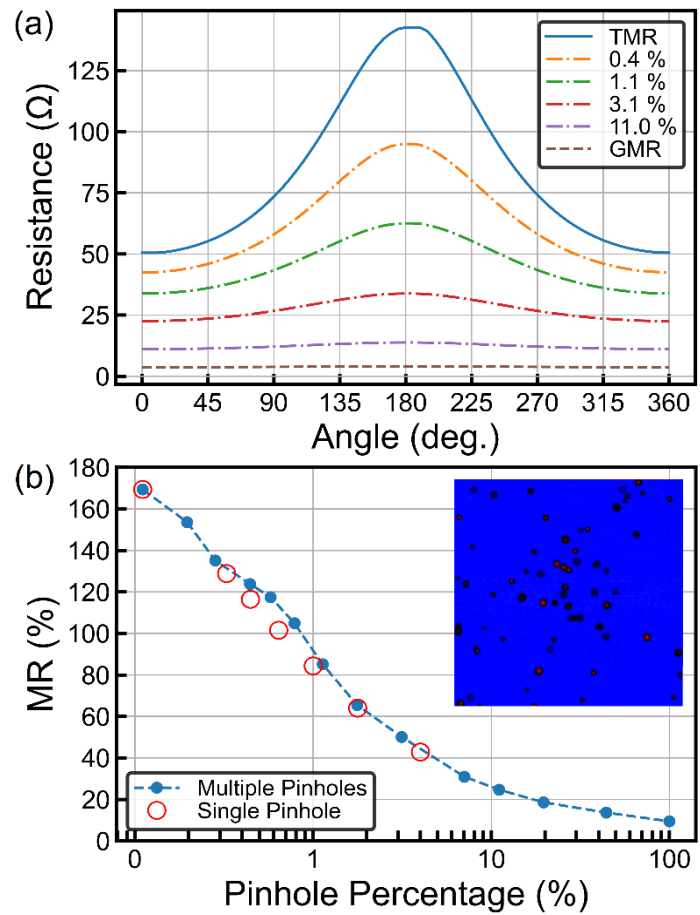

Next, the effect of metallic pinhole defects in the insulator layer is analysed, defined as cylindrical metallic conduction channels, with diameter in the range 1 – 2 nm, and conductivity varying with a tanh wall profile from 0 at the border to 5 MS/m at the centre. A typical randomly generated pinhole defect configuration is depicted in the inset to Figure 3(b), with blue signifying tunnelling conductance, and red signifying metallic conductance. Two cases are analysed: i) fixed MTJ square cross-sectional area of 40 nm × 40 nm and multiple pinholes, ii) a single pinhole, but varying square cross-sectional area from 40 nm × 40 nm down to 10 nm × 10 nm. In both cases we define the pinhole percentage as the percentage area covered by pinholes. As a limiting case a fully metallic spin valve has also been simulated. Results are shown in Figure 3(a), plotting the computed resistance as a function of angle. The resistance was computed self-consistently through the entire N/F/I/F/N stack, obtained as the potential drop divided by the total charge current normal to a designated ground electrode. For the ideal MTJ case, the TMR value is ~180%, close to the set theoretical limit of 200%, lowered by the additional fixed resistance contributions from the N and F layers. As the pinhole percentage is increased, the TMR value rapidly degrades, following an inverse dependence as shown in Figure 3(b), both for the single and multiple pinholes cases. In the fully metallic transport limit an MR value of ~10% is obtained due to CPP-GMR.

Using the same pinhole distributions, the angular dependence of spin torques is also computed. The DL-STT is shown in Figure 4(a), normalized to the charge current density, for the ideal MTJ and fully metallic spin valve cases, and also showing results for mixed conductance cases with selected pinhole percentages. Whilst for the ideal MTJ case the STT follows the expected sinusoidal dependence, increasing the preponderance of metallic conductance channels degrades the STT polarization, and gradually increases the STT skew, as expected for metallic spin valves. The full results are shown in Figure 4(b), plotting the STT polarization and STT skew as a function of pinhole percentage. The results obtained here are important for MTJs with small cross-sectional area, particularly as this is reduced towards the 10 nm node as required for commercial applications, including read-heads [1] and STT-MRAM [3]. The results in Figure 3(b) and Figure 4(b) show that even a single pinhole defect can have a significant effect, particularly on TMR degradation, but also due to reduced STT polarization. The model developed here can also be used to compute magnetization dynamics in MTJs, by coupling the self-consistent spin transport solver to the magnetization dynamics equation, e.g. as done in Ref. [39]. Investigation of magnetization dynamics in MTJs however is left for a future work.

**Figure 4** – Spin torques in the N/F/I/F/N stack with metallic pinhole defects, computed self-consistently using the spin transport solver. (a) DL-STT normalized to the charge current density in the anti-parallel configuration, $J_c$, as a function of angle, shown for varying pinhole percentage, including the ideal case with no defects (Tunnelling). As a limiting case, the result obtained for a metallic spin valve is also shown (Metallic). (b) STT polarization and skew as a function of pinhole percentage. Solid symbols are for a fixed square cross-sectional area of 40 nm × 40 nm and with multiple pinholes, whilst open symbols are for a single pinhole but decreasing cross-sectional area.

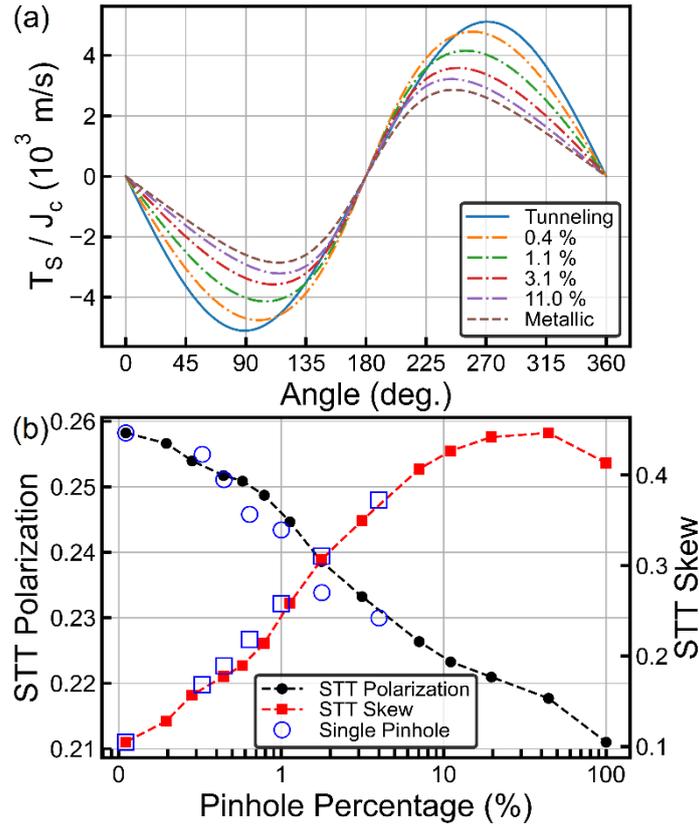

## IV. HDD Read-Heads

Using the model developed here, we now analyse a realistic HDD TMR read-head, in order to compute the signal-to-noise ratio (SNR), and the effect of bias voltage / current density on it. A generic stack of an HDD reader is shown in Figure 5(a).

**Figure 5** – Noise spectrum of an HDD reader at 300 K, with a reading voltage applied. (a) HDD reader geometry, (b) noise spectrum at room temperature, showing the FMR peaks of FL, RL and PL. (c) Detail of the noise spectrum, as a function of reading voltage, showing increase in $1/f$ noise with increasing voltage, where the operating frequency range is identified between 50 MHz and 2 GHz with vertical dashed lines.

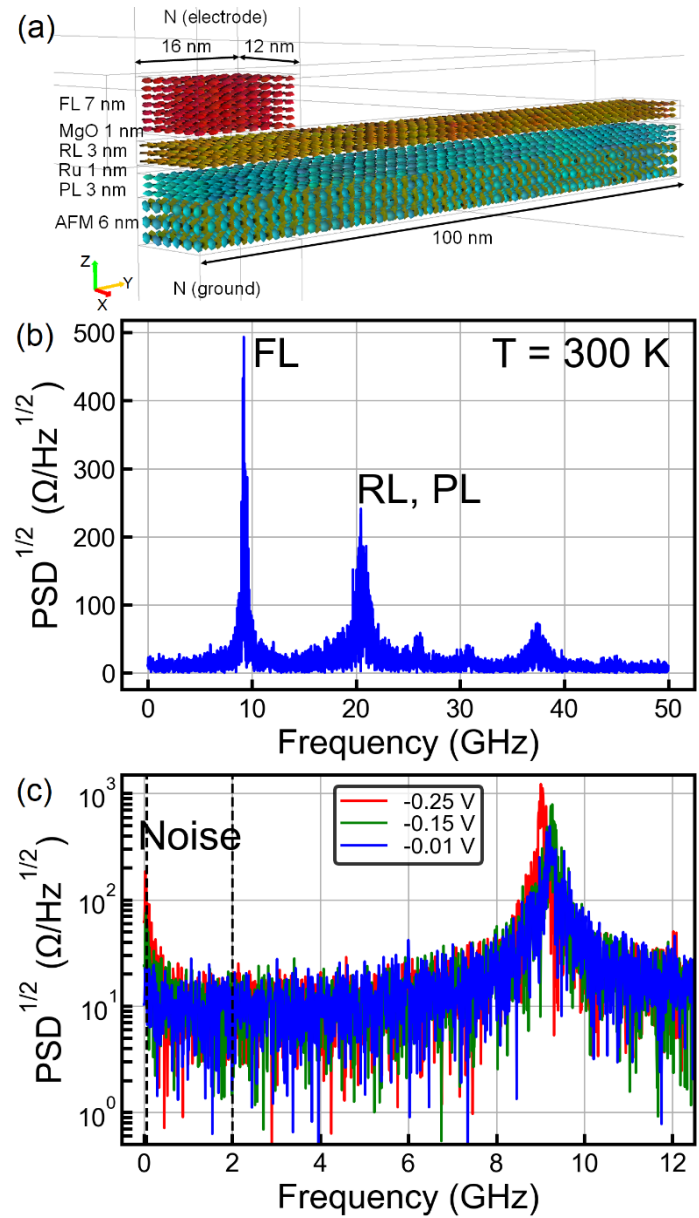

Here, the free layer (FL) is biased by the side shields (not shown on the figure), approximated by permanent magnets with $\mu_0 M_s$ = 1.5 T, and isolation thickness between FL and side shields from each side in the $x$ direction of 4 nm. A synthetic antiferromagnet (SAF) is comprised of the reference layer (RL), exchange coupled to the pinned layer (PL) via Ruderman–Kittel–Kasuya–Yosida (RKKY) interactions in a 0.8nm thick Ru spacer layer. The PL is stabilized by exchange coupling to an underlying antiferromagnetic (AFM) layer. Simulations are performed using the stochastic LLG equation at 300 K, with a cellsize of 2 nm × 2 nm × 1 nm, and integrated using the RK4 method [40] with a 50 fs time-step. The magnetic stray fields between the layers are fully included in the computation using the multi-layered convolution algorithm previously developed [41]. Further details, and all material parameters used for modelling the read-head of Figure 5(a), are given in Appendix C.

A typical reader's FMR spectrum is shown in Figure 5(b), obtained by a Fourier transform of the time-dependent device resistance over 100 ns, saved at fixed intervals of $\Delta t_s$ = 10 ps. The first peak observed in the spectrum (~9 GHz) corresponds to the ferromagnetic resonance (FMR) frequency of the FL, whilst the second peak (~20 GHz) corresponds to the FMR frequency of the RL and PL. The operating frequency range of a reader, used for noise integration, is in the range 50 MHz – 2 GHz, as indicated in Figure 5(c). In this range an equivalent noise resistance, $R_{noise}$, is obtained and magnetic noise SNR is defined as:

$$SNR(dB) = 20\log_{10}(\Delta R / R_{noise}) . \tag{11}$$

Here $\Delta R$ is the signal resistance amplitude between maximum deflection points of the FL for data bits 0 and 1 – details of $\Delta R$ computation are given in Appendix C.

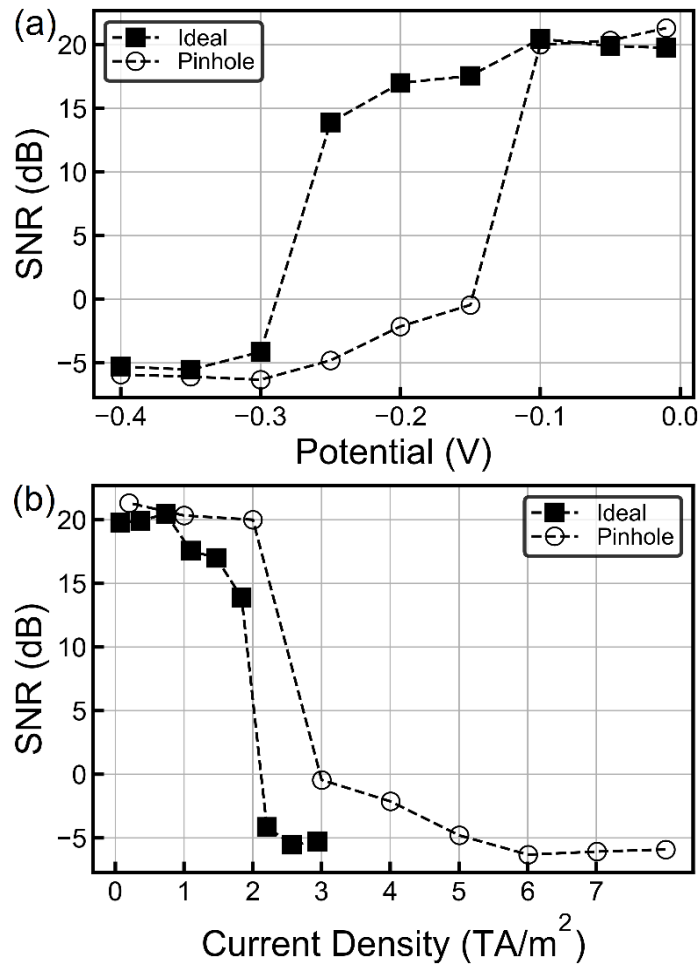

**Figure 6** – Signal to noise ratio in a magnetic read-head, with a metallic pinhole in the insulator layer, as well as without a pinhole (ideal). (a) SNR plotted as a function of potential, and (b) SNR plotted as a function of average current density. A single metallic pinhole has been used, resulting in ~1% pinhole area coverage.

It is known that spin torques in magnetic read-heads can induced instabilities [7,8], resulting in increased magnetic noise. Here we investigate this in the structure shown in Figure 5(a), by applying a potential across the read-head stack, and computing the spin torques in the entire stack self-consistently throughout the full simulated run of 100 ns. With a negative potential, the current density is along the $+z$ direction, and the spin torques between RL and PL are of damping type (in addition to the field-like component), which has the effect of stabilizing RL. Conversely, a positive potential results in anti-damping spin torques between RL and PL, which induces increased undesired fluctuations of the RL, thus here we restrict the analysis to the negative potential range. In the macrospin approximation, the spin torque on the FL is still of the form given in Equation (6), however now the STT polarization and field-like coefficients depend not only on RL, but also on PL. This is because RL is too thin to fully absorb the transverse spin components from PL, and the spin accumulation in the entire structure needs to be considered – i.e. the non-local character of spin torques in such a device with thin layers is important. A detailed quantitative comparison between the macrospin approximation and self-consistent spin torque approach for magnetic read-heads is outside the scope of the current work. However, we note from Equation (6) that thermal fluctuations in the RL (**p**) result in spin torque fluctuations on FL. This acts as a stimulus to further excite FMR oscillations in the FL, which increases as the potential is increased. Thus, on the one hand the FL resonance peak increases in height, but also the increased fluctuations give rise to increased $1/f$ noise, as may be seen in Figure 5(c).

The SNR as a function of potential is plotted in Figure 6(a), where we first analyse an ideal reader (i.e. without a pinhole defect in the tunnel barrier). As the potential is increased, fluctuations in FL become significant, resulting in a gradual decrease of the SNR. As the potential is further increased, a threshold is reached above which the fluctuations of FL are very large, greatly exceeding the signal amplitude $\Delta R$. This sets a maximum operating reading voltage range. We further consider the effect of a single metallic pinhole in the tunnel barrier, as in the previous section – results are plotted in Figure 6. At low voltages the SNR is observed to be comparable to the ideal read-head. The effect of the pinhole is to reduce the MR (see Figure 3), thus resulting in decreased signal amplitude, however, the noise is also reduced in proportion. The most important effect of the pinhole on the read-head is a reduction of the stability operating range. With a pinhole present, the operating reading voltage range is severely limited, since the current density, and hence the spin torque, is much greater for the same potential. When the SNR is plotted as a function of current density – Figure 6(b) – it is observed that the current density instability threshold is larger for the read-head with a metallic pinhole, compared to the ideal read-head. This is consistent with the results of the previous section – Figure 4 – where it was found that the STT efficiency decreases as metallic pinholes are introduced.

# V. Conclusions

Here, a three-dimensional drift-diffusion model was extended to include tunnel junctions, allowing computations in structures which include both tunnelling and metallic conduction. This approach allows self-consistent computation of spin torques, as well as magneto-resistance, which can include both TMR and CPP-GMR contributions. This is important for devices with non-uniform currents, such as tunnel barriers which include metallic pinhole defects, as well as devices with multiple magnetic layers, such as an HDD read-head, where the non-local character of spin torques needs to be taken into account. For a single tunnel junction separating two ferromagnetic layers, it is possible to solve the drift-diffusion model to a good approximation, obtaining the STT polarization and field-like coefficient parameters. Whilst for thick ferromagnetic layers the solution is exact, as the layer thicknesses are decreased, additional diffusive contributions arise at the ferromagnetic-metallic backing contact interfaces, due to spin-dependent scattering of incompletely absorbed transverse spin components. This results in a decrease of the STT polarization, dependent on both layers either side of the barrier. In this limit the thick ferromagnetic layer approximation breaks down, and instead we propose an orthogonal approximation, which takes into account these additional diffusive contributions. In general however, the fully self-consistent numerical approach is more powerful, and we analysed the effect of metallic pinholes in a tunnel barrier, both for a single MTJ, as well as a realistic read-head geometry comprising multiple magnetic layers. Increasing the area covered by pinholes results in a rapid degradation of the magneto-resistance, following an inverse dependence. Moreover, the spin torque angular dependence becomes skewed, similar to that obtained in fully metallic spin valves, and the STT polarization decreases. For HDD readers, the presence of even a single metallic pinhole severely reduces the region of reading voltage operating stability, as STT-induced magnetic fluctuations decrease the SNR below 0 dB. The model introduced here is useful in general for computation of spin torques in more complex multi-layered stacks with realistic layer thicknesses, and in particular in HDD read-heads, where the non-local nature of spin torques, as arising from spin-dependent scattering at multiple interfaces, needs to be taken into account.

# Appendix A

Material parameters used for simulating the N/F/I/F/N stack are given in Table 1. Equation (4) was solved using the successive over-relaxation (SOR) method, with a convergence error of $10^{-7}$ and centred finite-difference cellsize of 0.5 nm, alternately for $V$ and $\mathbf{S}$ until both Poisson equations are solved to the set convergence error. After each SOR iteration, values in cells at interfaces between materials are computed by taking currents in Equation (1) to be continuous normal to the interface, to second order accuracy in space. Simulations were performed using BORIS [42].

**Table 1** – Material parameters used for spin transport simulations in the N/F/I/F/N stack.

| \ | \ |
|---|---|
| **Metallic Contacts – N** | |
| $D_{e,N}$ | $4.0 \times 10^{-3}$ m$^2$/s |
| $\sigma_N$ | $20 \times 10^6$ S/m |
| $\lambda_{sd}$ | $5.0 \times 10^{-9}$ m |
| **Ferromagnets – F** | |
| $D_{e,F}$ | $1.0 \times 10^{-3}$ m$^2$/s |
| $\sigma_F$ | $4.0 \times 10^6$ S/m |
| $\lambda_{sf}$ | $5.0 \times 10^{-9}$ m |
| $\lambda_J$ | $2.0 \times 10^{-9}$ m |
| $\lambda_\varphi$ | $4.0 \times 10^{-9}$ m |
| $\beta_D$ | 0.8 |
| $P$ | 0.4 |
| **Insulator (Tunnelling) – I** | |
| $R_p A$ | $75 \times 10^{-15}$ $\Omega$m$^2$ |
| $R_{ap} A$ | $225 \times 10^{-15}$ $\Omega$m$^2$ |
| **Insulator (Metallic Pinholes)** | |
| $D_e$ | $1.0 \times 10^{-3}$ m$^2$/s |
| $\sigma$ | $5 \times 10^6$ S/m |
| $\lambda_{sd}$ | $5.0 \times 10^{-9}$ m |

It should be noted the parameters in Table 1 are readily accessible by experiment. In addition to the electrical conductivity, electron diffusion constants, and the tunnel barrier *RA* product, the spin diffusion length, $\lambda_{sf}$, and current spin polarization, *P*, may be also experimentally determined, for example using spin-absorption technique in lateral spin valves [43]; for a review of methods from an experimental point of view see Ref. [44]. The diffusion spin polarization, $\beta_D$, is well known from CPP-GMR studies, and has been highly successful in reproducing CPP-GMR experimental results [22]; also see Ref. [44]. The exchange rotation and spin dephasing lengths, $\lambda_J$ and $\lambda_\varphi$ respectively, may also be obtained, either from *ab initio* modelling, or by reproducing CPP-GMR experimental results as previously shown [31].

## Appendix B

The general spin accumulation solution in a N/F/I/F/N stack requires 18 boundary values, obtained using 18 equations based on the continuity of **S** and **J**$_{Sz}$ components (6 equations for each of the N/F, F/F and F/N interfaces – here the I layer is treated as having zero thickness since the net spin current is conserved across it). The solution in Equation (**8**) is not the general solution, but is only applicable to the case of thick F layers, where we can effectively decouple the effect of the N layers. The general solution in the right F layer is given as (*z* between 0 and *d$_F$*):

$$S_x(z) = \exp(-k_1 z)\left[u \sin(k_2 z) + v \cos(k_2 z)\right] + G \exp((z - d_F)/\lambda_{sf})$$
$$S_y(z) = S_F \exp(-z/\lambda_{sf}) + S_N \exp((z - d_F)/\lambda_{sf}), \quad (12)$$
$$S_z(z) = \exp(-k_1 z)\left[v \sin(k_2 z) - u \cos(k_2 z)\right] + H \exp((z - d_F)/\lambda_{sf}).$$

For the thick F layer approximation the exponential terms with leading coefficients *G* and *H* are negligible, which results in a significant simplification. For thin F layers however, where the transverse spin components are not fully absorbed, the transverse spin currents incident on the backing N layer give rise to the additional exponential terms in Equation (12), which are additional diffusive contributions due to spin-dependent scattering. Thus, the general solution in the backing N layer becomes (*z* > *d$_F$*):

$$S_x(z) = S_{x0} \exp(-(z - d_F)/\lambda_{sd}),$$
$$S_y(z) = S_{y0} \exp(-(z - d_F)/\lambda_{sd}), \quad (13)$$
$$S_z(z) = S_{z0} \exp(-(z - d_F)/\lambda_{sd}).$$

Equations (**12**) and (**13**) contain 9 boundary values (*u*, *v*, *G*, *H*, *S$_F$*, *S$_N$*, *S$_{x0}$*, *S$_{y0}$*, *S$_{z0}$*) which need to be determined (hence 18 boundary values for the entire N/F/I/F/N stack for the general asymmetric case which we solve here – the simpler symmetric case, i.e. reference and free layers of same thickness, only requires 9 boundary values overall).

The difficulty here is $k_1$ and $k_2$ cannot be decoupled from the boundary values as for the thick F layer approximation. Thus, substituting Equation (12) in Equation (4) with uniform magnetization, the following relation is obtained:

$$
\begin{aligned}
&[(k_1^2 - k_2^2 - \lambda_{sf}^{-2} - \lambda_{\varphi}^{-2})u + (2k_1 k_2 - \lambda_J^{-2})v]\sin(k_2 z) = \\
&[-(k_1^2 - k_2^2 - \lambda_{sf}^{-2} - \lambda_{\varphi}^{-2})v + (2k_1 k_2 - \lambda_J^{-2})u]\cos(k_2 z) \\
&+ \lambda_J^{-2} H \exp((z - d_F)/\lambda_{sf} + k_1 z) + \lambda_{\varphi}^{-2} G \exp((z - d_F)/\lambda_{sf} + k_1 z)
\end{aligned}
\quad (14)
$$

In the absence of exponential decay terms, sin and cos functions may be alternately integrated out using their orthogonality property, which allows elimination of both $u$ and $v$, obtaining the simple expressions $(k_1 \pm ik_2)^2 = \lambda_{sf}^{-2} + \lambda_{\varphi}^{-2} \pm i\lambda_J^{-2}$. If the exponential terms are not negligible, as is the case for thin F layers, then the left and right sides of Equation (14) are no longer orthogonal, which means $k_1$ and $k_2$ now depend on $u$, $v$, $G$, and $H$, making the overall exact solution an implicit problem, requiring a numerical method to solve. However, we can take the left and right sides of Equation (14) as being approximately orthogonal to obtain the above explicit formulas for $k_1$ and $k_2$, but the $G$ and $H$ values are now taken as non-zero. Expressions for $S_{x0}$, $S_{y0}$, $S_{z0}$ are obtained from continuity of spin accumulation at the N/F interfaces. The remaining 12 boundary values are obtained by inverting a 12×12 matrix, based on the remaining continuity relations at interfaces. Expressions are not reproduced here as they are very lengthy, however the process is relatively straightforward.

In order to investigate the range of validity of the orthogonal approximation, results are shown in Figure 7 as a function of reference and free layer thickness. With a fixed reference layer thickness of 5 nm, it may be seen in Figure 7(a) the numerical results are closely reproduced by the orthogonal approximation solution, but not by the thick F layer approximation. Spin-dependent scattering of transverse spin components at the N/F interface of the reference layer reduces the net spin current in both the reference and free layers, and thus the spin torques in the free layer also depend on the reference layer thickness. This is not reproduced by the thick F layer approximation, which doesn't take into account the effect of a finite reference layer thickness. In Figure 7(b) we also plot results for a fixed free layer thickness of 5 nm, as a function of reference layer thickness, showing the reduction in both the DL-STT and FL-STT as the reference layer thickness decreases. The orthogonal approximation solution is seen to breakdown for thin reference layers below 5 nm thickness, where a numerical solution is required to compute the spin torque parameters accurately.

**Figure 7** – Spin torques in an ideal tunnel junction, computed self-consistently using the spin transport solver and compared to analytical solutions. (a) Fixed reference layer (RL) thickness of 5 nm and varying free layer (FL) thickness. The numerical results are compared to the analytical solution from Equation (9) – the thick F layer approximation – and the orthogonal approximation discussed here. (b) Fixed FL thickness of 5 nm and varying RL thickness. The numerical results are compared to the analytical solution using the orthogonal approximation, which is seen to breakdown for RL below 5 nm thickness. The thick F layer approximation incorrectly obtains results independent of RL thickness.

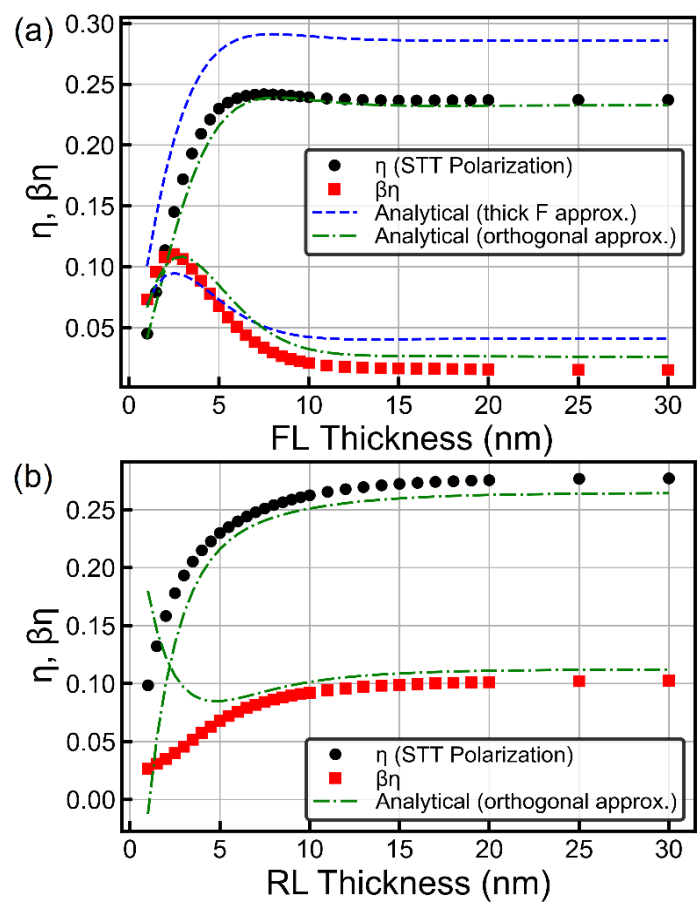

# Appendix C

In addition to the transport parameters shown in Appendix A, the magnetic material parameters used to model the magnetic read-head are shown in Table 2 ($M_S$: saturation magnetization, $K_1$: 2$^{nd}$ order uniaxial magneto-crystalline anisotropy constant, **e**: easy axis direction, $A$: exchange stiffness, $\alpha$: damping, $J_1$: bilinear surface exchange – BSE). Additionally, the AFM layer has a homogeneous antiferromagnetic exchange constant of -10$^6$ J/m$^3$. All the equations used to model the reader in BORIS (stochastic LLG, both one and two-sublattice, exchange, magneto-crystalline anisotropy, demagnetizing field, RKKY, exchange bias – EB) are given in Ref. [42].

**Table 2** – Magnetic material parameters used for modelling the magnetic read-head.

|        | $M_S$ (A/m) | $K_1$ (J/m$^3$) | **e** (**x, y, z**) | $A$ (J/m)      | $J_1$ (J/m$^2$)         | $\alpha$ |
|--------|-------------|-----------------|---------------------|----------------|-------------------------|----------|
| **FL** | 1150×10$^3$ | 3.6×10$^3$      | (1, 0, 0)           | 21×10$^{-12}$  | 0.1×10$^{-3}$ (BSE)     | 0.01     |
| **RL, PL** | 1430×10$^3$ | 36×10$^3$   | (0, 1, 0)           | 21×10$^{-12}$  | -1.5×10$^{-3}$ (RKKY)   | 0.01     |
| **AFM** | 800×10$^3$ | 100×10$^3$     | (0, 1, 0)           | 13×10$^{-12}$  | 1.0×10$^{-3}$ (EB)      | 0.02     |

The resistance signal amplitude is calculated for data bits with 50 nm × 20 nm dimensions, 10 nm thickness, and 6 nm reader-to-bit separation. The saturation magnetization of data bits is set to 800 kA/m, and bits 0 and 1 are defined as magnetization along –y and +y directions respectively. The difference in the two resistance states, obtained for relaxed magnetization configurations, gives the resistance signal amplitude, $\Delta R$. This is plotted in Figure 8 as a function of reading potential, both for an ideal read-head, and one with a metallic pinhole defect. As the potential is increased, spin torques between the layers, and in particular between FL and RL, result in a shifting of the FL and RL equilibrium directions, resulting in a change of $\Delta R$ with potential. For negative potential values, the STT acts to shift the FL and RL magnetization orientations towards the anti-parallel orientation, obtaining a monotonic decrease in resistance for both bits 0 and 1. This shift is asymmetric however for the two bits, and thus $\Delta R$ does not decrease monotonically with potential. In particular, it may be seen this effect is non-linear at higher current densities, as seen in Figure 8(b) for the HDD reader with a pinhole defect, where the current density range is greater for the same potential range.

**Figure 8** – Resistance signal amplitude for the magnetic read-head of Figure 5, obtained using a data bit with 50 nm × 20 nm dimensions, 10 nm thickness, and 6 nm reader-to-bit separation, as a function of the bias voltage. The data bit saturation magnetization is set to 800 kA/m, and the signal amplitude Δ$R$ is obtained as the difference in resistance states of the read-head, for bit magnetization directions –$y$ and +$y$ respectively. The resistance signal amplitude is shown for (a) an ideal read-head (no pinhole defects), and (b) a single metallic pinhole defect.

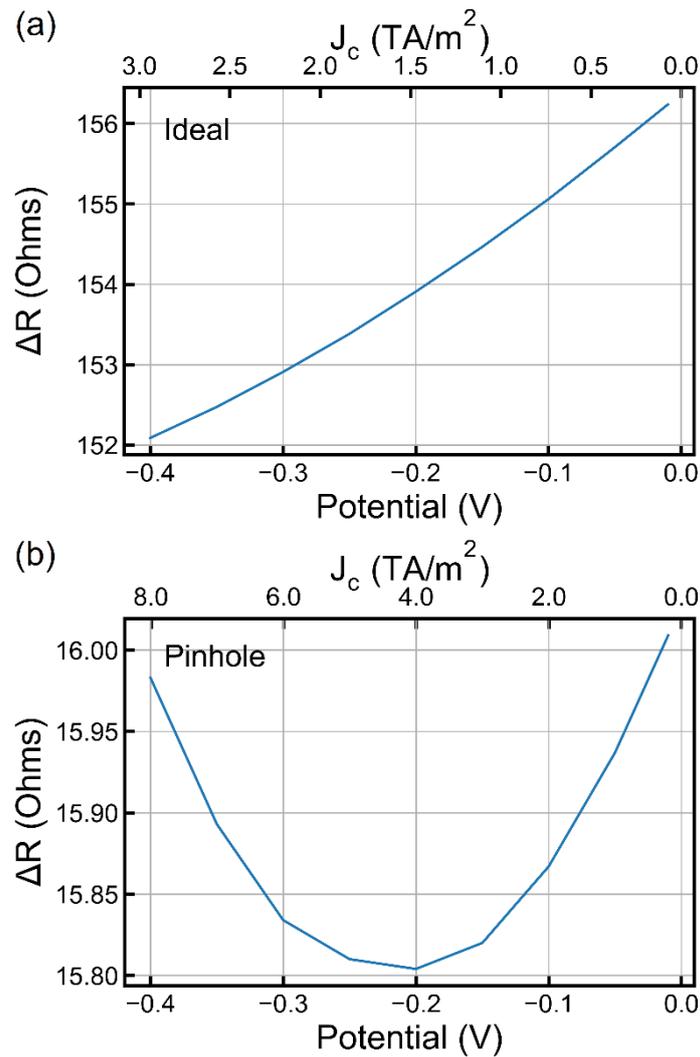